\documentstyle[epsfig]{l-school}
\begin{document}
\newcommand{\be}{\begin{eqnarray}}
\newcommand{\ee}{\end{eqnarray}}
\newcommand\del{\partial}
\newcommand\barr{|}
\newcommand\half{\frac 12}
\markboth{J.J.M. Verbaarschot}{SPECTRAL FLUCTUATIONS}
%
\setcounter{part}{1}
%
\title{Spectral Fluctuations of the QCD Dirac Operator}
%
\author{J.J.M. Verbaarschot}
%
\institute{Department of Physics,\\
SUNY at Stony Brook,\\
Stony Brook, NY 11794, USA}
\maketitle
\section{INTRODUCTION}

It is generally believed that QCD with massless quarks undergoes a chiral
phase transition (see \cite{DeTar} for a review). This
leads to important observable signatures in
the real world with two light quarks. The order parameter of the chiral phase
transition is the chiral condensate $\langle \bar \psi \psi \rangle$, 
is directly related to the average spectral density  
of the Dirac operator \cite{Banks-Casher}.  
However, the eigenvalues of the Dirac operator fluctuate about their average
position. The question we wish to address in this lecture is to what
extent such fluctuations are universal. If that is the case, they 
do not depend on the full QCD dynamics and can
be obtained from a much simpler chiral Random Matrix Theory (chRMT) 
with the global symmetries of the QCD partition function. 

This conjecture has its origin in the study of 
spectra of complex systems \cite{bohigas}. According to the Bohigas
conjecture, spectral correlations of classically chaotic quantum systems
are given by RMT. A first argument in favor of 
the universality in Dirac spectra
came from the analysis of the finite volume QCD partition function.
In particular, for box size $L$ in the range 
$
1/\Lambda \ll L \ll 1/m_\pi,
$
($\Lambda$ is a typical hadronic scale and $m_\pi$ is the pion mass)
we expect that the global symmetries determine its mass dependence
\cite{LS}. This implies that the fluctuations of 
Dirac spectra are constrained by Leutwyler-Smilga sum rules.

Recently, it has become possible to obtain $all$ eigenvalues of the lattice
QCD Dirac operator on reasonably large lattices \cite{Kalkreuter}. This makes
a direct verification of the above conjecture possible. This is the main 
objective of this lecture.

At nonzero chemical potential the QCD Dirac operator is nonhermitean with
eigenvalues scattered in the complex plane. The possibility of 
a new type of universal behavior
in this case will be discussed at the end of this lecture. 

 
\section{THE DIRAC SPECTRUM}
The order parameter of the chiral phase transition,
$\langle \bar \psi \psi \rangle$,
is nonzero only below the critical temperature. 
As was shown in \cite{Banks-Casher} 
$\langle \bar \psi \psi \rangle$ is directly related to the eigenvalue density
of the QCD Dirac operator per unit four-volume 
\be
\Sigma \equiv 
|\langle \bar \psi \psi \rangle| =\frac {\pi \langle {\rho(0)}\rangle}V.
\label{bankscasher}
\ee
It is elementary to derive this relation.
The Euclidean Dirac operator for 
gauge field configuration $A_\mu$ is given by
$
D = \gamma_\mu (\del_\mu + i A_\mu).
$
For Hermitean gamma matrices 
$D$ is anti-hermitean with purely imaginary eigenvalues,
$
D\phi_k = i\lambda_k \phi_k,
$
and  spectral density given by
$
\rho(\lambda) = \sum_k \delta (\lambda - \lambda_k).
$
Because $\{\gamma_5, D\} = 0$, 
nonzero eigenvalues occur in pairs $\pm \lambda_k$. 
In terms of the eigenvalues of $D$ the QCD partition function
for $N_f$ flavors of mass $m$ can then be written as
\be
Z(m) = \langle \prod_k(\lambda_k^2 + m^2)^{N_f} \exp(-S_{\rm YM})\rangle,
\label{part}
\ee
where the average $\langle \cdot \rangle$ 
is over all gauge field configurations.

The chiral condensate follows immediately from
the partition function (\ref{part}),
\be
\langle \bar \psi \psi \rangle  = \frac 1{VN_f}\del_m \log Z(m) 
=\frac 1V \langle \sum_k \frac {2m}{\lambda_k^2 + m^2}\rangle.
\ee
If we express the sum as an integral over the spectral density,
and take the thermodynamic limit before the chiral limit so that we have many
eigenvalues less than $m$ we recover (\ref{bankscasher}) (Notice
the order of the limits.).

An important consequence of the Bank-Casher formula (\ref{bankscasher})
is that the eigenvalues near zero virtuality are spaced as  
$
\Delta \lambda = 1/{\rho(0)} = {\pi}/{\Sigma V}.
$
This should be contrasted with the eigenvalue spectrum 
of the non-interacting 
Dirac operator. Then
$
\rho^{\rm free}(\lambda) \sim V\lambda^3
$
which leads to an eigenvalue spacing of $\Delta \lambda \sim 1/V^{1/4}$.
Clearly, the presence of gauge fields leads to a strong modification of
the spectrum near zero virtuality. Strong interactions result in the 
coupling of many degrees of freedom leading to extended states and correlated
eigenvalues.
On the other hand, for uncorrelated eigenvalues, the eigenvalue distribution
factorizes and we have $\rho(\lambda) \sim \lambda^{2N_f+1}$, i.e. no breaking
of chiral symmetry. 

Because the
QCD Dirac spectrum is symmetric about zero,
we have two different types of eigenvalue correlations:
correlations in the bulk of the spectrum and spectral correlations near zero 
virtuality. In the context of
chiral symmetry we wish to study 
the spectral density near zero virtuality. Because the
eigenvalues are spaced as $1/\Sigma V$ it is natural to introduce the
microscopic spectral density \cite{SVR}
\be
\rho_S(u) = \lim_{V\rightarrow \infty} \frac 1{V\Sigma}
\rho\left( \frac u{V\Sigma} \right).
\ee
The dependence on the macroscopic variable $\Sigma$  has been eliminated
and therefore $\rho_S(u)$ is a perfect candidate for a universal function.


\section{SPECTRAL UNIVERSALITY}
Spectra for a wide range of complex quantum systems 
have been studied both experimentally and numerically
(see \cite{GMW} for a review). 
One basic observation
has been that the scale of variations of the average spectral
density and the scale of the spectral fluctuations separate. 
This allows us to unfold the spectrum, i.e. we rescale the 
spectrum in units of the local average level spacing. The fluctuations of the
unfolded spectrum can be measured by suitable statistics. We will consider the
nearest neighbor spacing distribution, $P(S)$, the
number variance, $\Sigma_2(n)$, and the $\Delta_3(n)$ statistic. The number
variance is defined as the variance of the number of levels in a stretch of
the spectrum that contains $n$ levels on average,
and $\Delta_3(n)$ is obtained by a smoothening of $\Sigma_2(n)$.

These statistics can be obtained analytically for the invariant random matrix
ensembles 
defined as ensembles of Hermitean 
matrices with independently distributed Gaussian matrix elements. 
Depending on the anti-unitary symmetry, the matrix elements are real, complex
or quaternion real. The corresponding Dyson index is given by
$\beta = 1,\, 2 $, and $4$, respectively. 
The nearest neighbor spacing distribution is given by 
$P(S) \sim S^{\beta}\exp(-a_\beta S^2)$. 
The asymptotic behavior of $\Sigma_2(n)$
and $\Delta_3(n)$ is
given by $\Sigma_2(n) \sim (2/\pi^2\beta) \log(n)$
and $\Delta_3(n) \sim \beta \Sigma_2(n)/2$. For uncorrelated eigenvalues
one finds that $P(S) = \exp(-S)$, $\Sigma_2(n) = n$ and $\Delta_3(n) = n/15$.
Characteristic features of random matrix correlations are
level repulsion at short distances and a strong suppression
of fluctuations at large distances.

The main conclusion of numerous studies of eigenvalue spectra is that
spectral correlations of a classically chaotic systems are given by RMT
\cite{GMW,hirsch}.

\section{CHIRAL RANDOM MATRIX THEORY}
In this section we will introduce an instanton liquid inspired 
RMT for the QCD partition function. 
In the spirit of the invariant random matrix ensembles 
we construct a model for the Dirac
operator with the global symmetries of the QCD partition function as input, but
otherwise  Gaussian random matrix elements. 
The chRMT that obeys these conditions is defined by
\cite{SVR,V,VZ}
\be
Z_\nu^\beta = \int DW \prod_{f= 1}^{N_f} \det({\rm \cal D} +m_f)
e^{-\frac{N\Sigma^2 \beta}4 {\rm Tr}W^\dagger W},\quad{\rm with}\quad
\label{zrandom}
{\cal D} = \left (\begin{array}{cc} 0 & iW\\
iW^\dagger & 0 \end{array} \right ),
\ee
and $W$ is a $n\times m$ matrix with $\nu = |n-m|$ and
$N= n+m$. The matrix elements of $W$ are either real ($\beta = 1$, chiral
Gaussian Orthogonal Ensemble (chGOE)), complex
($\beta = 2$, chiral Gaussian Unitary Ensemble (chGUE)),
or quaternion real ($\beta = 4$, chiral Gaussian Symplectic Ensemble (chGSE)).

This model reproduces the following symmetries of the QCD partition
function:
{\it i)} The $U_A(1)$ symmetry. All nonzero eigenvalues of the random matrix
Dirac operator occur in pairs $\pm \lambda$.
{\it ii)}  The topological structure of the QCD partition function. The 
Dirac matrix has exactly $|\nu|\equiv |n-m|$ zero eigenvalues. This identifies
$\nu$ as the topological sector of the model.
{\it iii)} The flavor symmetry is the same as in QCD.
{\it iv)} The chiral symmetry is broken spontaneously with 
chiral condensate given by
$                                                      
\Sigma = \lim_{N\rightarrow \infty} {\pi \rho(0)}/N.
$
($N$ is interpreted as the (dimensionless) volume of space
time.)
{\it v)} The anti-unitary symmetries. 
For  fundamental fermions the matrix elements of the Dirac operator are complex
for $N_c \ge 3$ ($\beta= 2$) but can be chosen real for $N_c = 2$ ($\beta =1$).
For adjoint fermions they can be arranged 
into real quaternions ($\beta = 4$).

Note that 
spectral correlations of chRMT in the bulk of the spectrum are given by the
invariant random matrix ensemble with the same value of $\beta$. 
Both microscopic correlations near zero virtuality and in the bulk of the
spectrum are 
stable against deformations of the ensemble. 
This has been shown by a variety of different arguments
\cite{zee,sener,Damgaard,tilo,sener2}.

Below we will discuss the microscopic spectral density. 
For $N_c = 3$, $N_f$ flavors and topological charge $\nu$
it is given by \cite{V}
\be
\rho_S(u) = \frac u2 \left ( J^2_{a}(u) -
J_{a+1}(u)J_{a-1}(u)\right),
\label{micro}
\ee
where $a = N_f + \nu$. The more complicated result for $N_c =2$
is given in \cite{V2}.

Together with the invariant random matrix ensembles, the chiral ensembles are
part of a larger classification scheme. As pointed out in \cite{class}, 
there
is a one to one correspondence between random matrix theories and symmetric
spaces.

\section{LATTICE QCD RESULTS}
Recently, Kalkreuter \cite{Kalkreuter}
calculated $all$ eigenvalues of the lattice Dirac
operator both for Kogut-Susskind (KS) fermions and Wilson fermions
for lattices 
as large as $12^4$. 
In the the case of $SU(2)$
the anti-unitary symmetry of the KS and the Wilson Dirac operator is
different  \cite{Teper,HV}. 
For KS fermions the
Dirac matrix can be arranged into real
quaternions, whereas 
the $Hermitean$ Wilson Dirac
matrix $\gamma_5 D^{\rm Wilson}$ can be chosen real.  
 Therefore, we expect that the eigenvalue
correlations are described by the GSE and the GOE, respectively \cite{HV}.
In Fig. 1 we show results for  $\Sigma_2(n)$,
$\Delta_3(n)$ and $P(S)$. 
The results for KS fermions are for 4 dynamical flavors
 with $ma = 0.05$ on a $12^4$ lattice. The results for Wilson fermion were
obtained for two dynamical flavors on a $8^3\times 12$ lattice.
Other statistics are discussed in \cite{HKV}.
\begin{center}
\begin{figure}[!ht]
\centering\includegraphics[width=95mm,angle=0]
{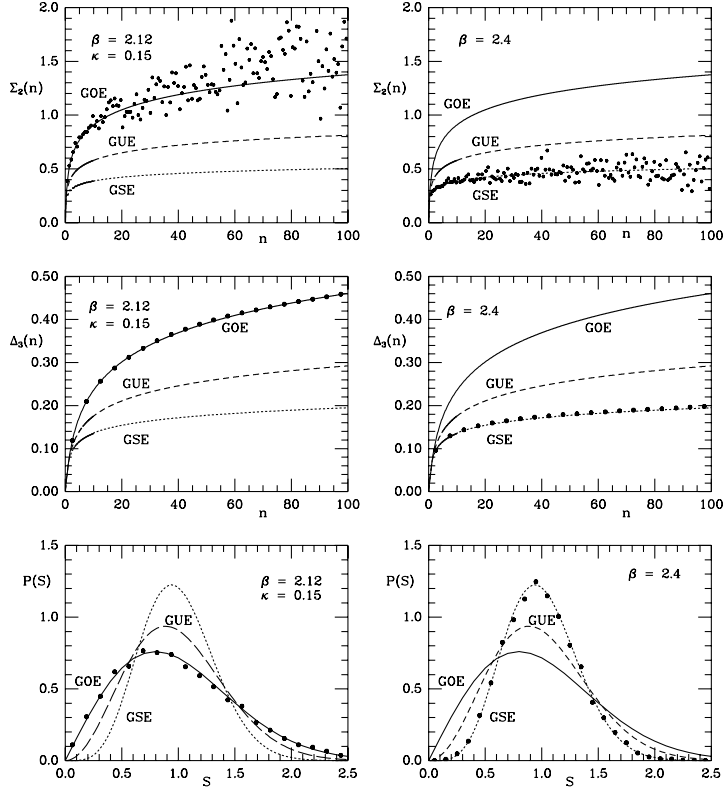}
\caption{
Spectral correlations of Dirac eigenvalues for Wilson fermions
(upper) and KS-fermions (lower). }
\label{fig1}
\vspace*{-0.5cm}
\end{figure}
\end{center}
\noindent

Recent lattice studies of the microscopic spectral density for quenched
$SU(2)$ show perfect agreement between the microscopic spectral density
and random matrix theory \cite{meyer}. These calculations were performed at
moderately strong coupling. Result for couplings in the scaling regime are
in progress. 

However,
an alternative way to probe the Dirac spectrum is via
the valence quark mass dependence of the condensate \cite{Christ},
i.e. $\Sigma(m) = \frac 1N \int d\lambda \rho(\lambda)
2m/(\lambda^2 +m^2)$, for a fixed sea quark mass. In the mesoscopic range,
$\Sigma(m)$ can be obtained analytically from
the microscopic spectral density (\ref{micro}) \cite{vplb},
\be
\frac {\Sigma(x)}{\Sigma} = x(I_{a}(x)K_{a}(x)
+I_{a+1}(x)K_{a-1}(x)),
\label{val}
\ee
where $x = mV\Sigma$ is the rescaled mass and $a = N_f+\nu$.
In Fig. 2 we plot this ratio as a function of $x$ for
lattice data of two dynamical flavors with  mass $ma = 0.01$ and $N_c= 3$ on a
$16^3 \times 4$ lattice.  We observe
that the lattice data for different values of $\beta$ fall on a single curve.
Moreover, in the mesoscopic range 
this curve coincides with the random matrix prediction for $N_f = \nu = 0$.
\begin{center}
\begin{figure}[!ht]
\centering\includegraphics[width=55mm,angle=270]{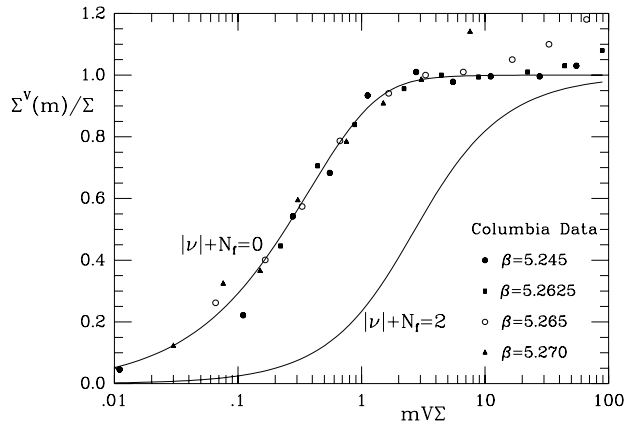}
\caption{
The valence quark mass dependence of the chiral condensate.}
\label{fig2}
\vspace*{-1.0 cm}
\end{figure}
\end{center}

\section{CHIRAL RANDOM MATRIX THEORY AT $\mu \ne  0$}
At nonzero temperature $T$ 
and chemical potential $\mu$ a $schematic$ random matrix 
matrix model of the Dirac operator in (\ref{zrandom}) is given by 
\cite{JV,Tilo,Stephanov}
\be
{\cal D} = \left (\begin{array}{cc} 0 & iW +i\Omega_T +\mu \\
iW^\dagger+i\Omega_T + \mu & 0 \end{array} \right ),
\label{diracmatter}
\ee
where $\Omega_T= T \otimes_n(2n+1)\pi {\large\bf 1}$.
Below, we will discuss a model with $\Omega_T$ 
absorbed in the random matrix and $\mu \ne 0$. 
For the three values of $\beta$ the eigenvalues of
${\cal D}$ are scattered in the complex plane. 

In the quenched approximation
the eigenvalue distribution for $\beta =2$ was obtained 
analytically \cite{Stephanov} from 
the $N_f\rightarrow 0$ limit of a partition function with the determinant 
replaced by its absolute value. The same analysis can be carried out
for $\beta =1$ and $\beta=4$. In the normalization (5), it turns out 
that the average spectral density does not depend on $\beta$ \cite{Osborn}. 
However, the 
fluctuations of the eigenvalues are $\beta$ dependent. 
In particular, one finds a very different behavior close to the imaginary
axis and not too large values of $\mu$. 
This regime of almost Hermitean
random matrices was first identified by Fyodorov et al. \cite{yan}.
For $\beta =1$ we observe an accumulation of purely imaginary eigenvalues,
whereas for $\beta =4$ we find a depletion of eigenvalues in this domain.
These results explain quenched instanton liquid calculations
\cite{Thomas} and lattice QCD simulations 
with Kogut-Susskind fermions \cite{baillie} at $\mu\ne 0$ 
(both for  $SU(2)$), respectively. 
\section{CONCLUSIONS}
We have shown that microscopic correlations of the QCD Dirac spectrum
can be explained by RMT and have obtained an analytical
understanding of the distribution of the eigenvalues near zero.
An extension of this model to nonzero chemical potential explains
previously obtained lattice QCD results. 
\ack{
This work was partially supported by the US DOE grant
DE-FG-88ER40388. M. Halasz and M. \c Sener
are thanked for a critical reading of the manuscript.}


\begin{thebibliography}{100}
\itemsep=0cm
\bibitem{DeTar}
C.~DeTar, {\it Quark-gluon plasma in numerical simulations of QCD}, in {\it
Quark gluon plasma 2}, R. Hwa ed., World Scientific 1995.
\bibitem{Banks-Casher}T.~Banks and A.~Casher, Nucl. Phys. {\bf B169} (1980) 103.
\bibitem{bohigas}O.~Bohigas, M.~Giannoni, Lecture notes in Physics
{\bf 209} (1984) 1.
\bibitem{LS}
H.~Leutwyler and A.~Smilga, Phys. Rev. {\bf D46} (1992) 5607.
\bibitem{Kalkreuter}T. Kalkreuter,  Comp. Phys. Comm. {\bf 95} (1996) 1.
\bibitem{SVR}E. Shuryak and J. Verbaarschot,
Nucl. Phys. {\bf A560} (1993) 306.
\bibitem{GMW}T. Guhr, A. M\"uller-Groeling and H.A. Weidenm\"uller, Phys. Rep.
(1997).
\bibitem{hirsch} J. Verbaarschot, Proceedings Hirschegg 1997, 
hep-ph/9705355.
\bibitem{V} J. Verbaarschot, Phys. Rev. Lett. {\bf 72} (1994) 2531; Phys. Lett.
{\bf B329} (1994) 351; Nucl. Phys. {\bf B427} (1994) 434.
\bibitem{VZ}J. Verbaarschot and I. Zahed,
Phys. Rev. Lett. {\bf 70} (1993) 3852.
\bibitem{zee}E. Br\'ezin, S. Hikami and A. Zee,
Nucl. Phys. {\bf B464} (1996) 411.
\bibitem{sener} A. Jackson, M. Sener and J. Verbaarschot, Nucl. Phys.
{\bf B479} (1996) 707.
\bibitem{Damgaard}S. Nishigaki, Phys. Lett. B (1996); G. Akemann, 
P. Damgaard, U. Magnea and S. Nishigaki, 
hep-th/9609174.
\bibitem{sener2} A. Jackson, M. Sener and J. Verbaarschot, 
hep-th/9704056.
\bibitem{tilo}T. Guhr and T. Wettig, hep-th/9704055.
\bibitem{V2}J. Verbaarschot, Nucl. Phys. {B426} (1994) 559.
\bibitem{class}M. Zirnbauer, J. Math. Phys. {\bf 37} (1996) 4986.
\bibitem{Teper}S. Hands and M. Teper, Nucl. Phys. {\bf B347} (1990)
819.
\bibitem{HV}M. Halasz and J. Verbaarschot,
Phys. Rev. Lett. {\bf 74} (1995) 3920.
\bibitem{HKV}M. Halasz, T. Kalkreuter and J. Verbaarschot, hep-lat/9607042.
\bibitem{meyer}  M.E. Berbenni-Bitsch et al., 
hep-lat/9704018.
\bibitem{Christ}S. Chandrasekharan, Lattice 1994, 475;
S. Chandrasekharan and N. Christ, Lattice 1995, 527; N. Christ, Lattice 1996.
\bibitem{vplb}J. Verbaarschot, Phys. Lett. {\bf B368} (1996) 137.
\bibitem{JV}A. Jackson and J. Verbaarschot, Phys. Rev. {\bf D53} (1996)
7223.
\bibitem{Tilo}T. Wettig, A. Sch\"afer and H. Weidenm\"uller,
Phys. Lett. {\bf B367} (1996) 28.
\bibitem{Osborn}M. Halasz, J. Osborn and J. Verbaarschot, hep-lat/9704007.
\bibitem{Stephanov}M. Stephanov, Phys.\ Rev.\ Lett.\ {\bf 76} (1996) 4472.
\bibitem{yan} Y. Fyodorov, B. Khoruzhenko and H. Sommers, Phys.
Lett. {\bf A 226}, 46 (1997); cond-mat/9703152.
\bibitem{Thomas} E. Shuryak and Th. Sch\"afer, {\it private communication}.
\bibitem{baillie} C. Baillie et al., Phys. Lett. {\bf 197B}, 195 (1987).
\end{thebibliography}
\end{document}